\documentclass[preprint,12pt,authoryear]{elsarticle}

\usepackage{amssymb}
\usepackage{amsthm}
\usepackage{amsmath}
\usepackage{booktabs}
\usepackage{numprint}
\usepackage{siunitx}
\usepackage{xurl}
\usepackage{xcolor}

\DeclareMathOperator{\Expect}{\mathbb{E}}
\DeclareMathOperator{\Var}{Var}

\begin{document}

\begin{frontmatter}

  \title{Bayesian modelling of herd-level infection dynamics in
    cattle: Local spread as the primary driver of \textit{Salmonella}
    Dublin persistence on Öland}

\author[1]{Stefan Widgren\corref{cor1}}
\ead{stefan.widgren@sva.se}

\author[1,2]{Ivana R. Ewerlöf}

\author[1]{Arianna Comin}

\author[1]{Robert Johansson}

\author[3,4]{Stefan Engblom}

\cortext[cor1]{Corresponding author}

\affiliation[1]{
  organization={Swedish Veterinary Agency (SVA), Department of
    Epidemiology, Disease Surveillance and Risk Assessment},
  city={Uppsala},
  postcode={SE-751 89},
  country={Sweden}}

\affiliation[2]{
  organization={Swedish University of Agricultural Sciences (SLU),
    Department of Applied Animal Science and Welfare},
  addressline={Box 234},
  city={Skara},
  postcode={SE-532 23},
  country={Sweden}}

\affiliation[3]{
  organization={Uppsala university, Department
    of Information Technology, Division of Scientific Computing},
  addressline={Box 337},
  city={Uppsala},
  postcode={SE-751 05},
  country={Sweden}}

\affiliation[4]{
  organization={Uppsala university, Science for Life Laboratory,
    Department of Information Technology},
  addressline={Box 337},
  city={Uppsala},
  postcode={SE-751 05},
  country={Sweden}}


\begin{abstract}
  \textit{Salmonella} Dublin (\textit{S.}~Dublin), a zoonotic serotype
  adapted to cattle, causes animal welfare issues and economic losses.
  The disease has proven particularly challenging to control in Öland,
  Sweden. This study uses Bayesian simulation-based inference of bulk
  tank milk sample results to analyse the \textit{S.}~Dublin infection
  dynamics in Öland cattle. The infection process was formulated as a
  dynamic state-space model and particle Markov-chain Monte Carlo
  methods were applied to infer the underlying infection dynamics and
  estimate the basic reproduction number ($R_0$) as well as the
  effective reproduction number ($R_t$).  These metrics provide
  insight into transmission dynamics, enabling assessment of the
  effectiveness of the current \textit{S.}~Dublin control in Swedish
  cattle and identification of interventions that may reduce the
  prevalence.

  The results show that most holdings on Öland have $R_0 < 1$,
  indicating that infection is expected to die out after introduction.
  However, in a subset of holdings $R_0 > 1$, and there the risk for
  spread of \textit{S.}~Dublin is higher. Furthermore, the analysis
  reveals that on average, $R_t \approx 1$, suggesting a stable
  endemic presence unless effective interventions are implemented. In
  addition, the results show that it is insufficient to restrict the
  movements of infected cattle on Öland to bring $R_t < 1$, as local
  spread and within-herd transmission contribute equally to the force
  of infection (approximately 50\% each). These findings demonstrate
  how Bayesian data-driven analysis can support evidence-based
  decision making for the control and eradication of
  \textit{S.}~Dublin in cattle.
\end{abstract}


\end{frontmatter}


\section{Introduction}
\label{introduction}

\textit{Salmonella} Dublin (\textit{S.}~Dublin) is a cattle-adapted
salmonella serotype with zoonotic potential \citep{Jones2008}.  It
causes clinical disease in cattle with symptoms such as fever,
diarrhoea, pneumonia, arthritis, abortion and can ultimately result in
mortality \citep{Nielsen2012, LaRagione2013}---and consequently leads
to reduced animal welfare and economic losses for the farmer
\citep{Nielsen2013a}.  \textit{S.}~Dublin infection in humans is rare,
but when it occurs, consequences are often more serious than those
caused by other serotypes \citep{Jones2008}.  In Europe,
\textit{S.}~Dublin is common in cattle, and some countries have
control programs to reduce the prevalence \citep{Bergevoet2009,
  Nielsen2013b, Agren2015}.

Since the 1960s, Sweden has maintained a comprehensive Salmonella
control program covering the entire chain from feed to food. This
multi-component approach, encompassing feed safety measures, strict
biosecurity protocols, and mandatory movement restrictions coupled
with on-farm eradication plans upon detection \citep{Agren2015}, has
successfully maintained Salmonella prevalence in cattle at a low and
steady level for the past three decades. However, with increased costs
associated with cattle-specific on-farm control measures, proposals
have been made to develop a more cost-efficient control program for
bovine salmonellosis in Sweden \citep{Vagsholm2007, SVA2025}. Several
bulk tank milk (BTM) screenings using enzyme-linked immunosorbent
assay (ELISA) for detecting antibodies against \textit{S.}~Dublin
indicate that approximately 1\% of dairy herds are currently
seropositive in Sweden overall, with substantial regional variation
ranging from 0\% in some regions to 15--20\% on Öland (an island in
the south-east) \citep{Nyman2013, Agren2016, Agren2017,
  SVA2024}. Similar regional differences in \textit{S.}~Dublin
prevalence have been observed in Dutch \citep{Fabri2024} and Danish
\citep{Ersboll2011} dairy herds. A better understanding of the drivers
behind these regional differences would facilitate the development of
risk-based surveillance and management strategies for
\textit{S.}~Dublin.

Several important risk factors have been identified for the spread of
\textit{S.}~Dublin between farms, including the introduction of
infected animals from other herds \citep{Vaessen1998, VanSchaik2002,
  Nielsen2007b, Conrady2024, Fabri2024}.  Additionally, local
transmission of \textit{S.}~Dublin between neighbouring farms is
another identified risk factor \citep{Nielsen2007b, Agren2016}.
However, the relative contribution of animal movements versus local
transmission to the overall force of infection remains uncertain.  No
consensus exists on whether herd size affects \textit{S.}~Dublin
infection and spread, where some studies find no association
\citep{Fossler2005a, Fossler2005b}, and others have shown that larger
herds are more at risk \citep{Vaessen1998, Nielsen2007b, Agren2016}.
It is possible, therefore, that the trend towards larger herds
\citep{Conrady2024, Ewerlof2025} may lead to an increase in the
prevalence of \textit{S.}~Dublin, but the relative importance of these
different risk factors needs to be further investigated.

We can gain valuable insights into the underlying infection process
within- and between herds by combining BTM sample results for
\textit{S.}~Dublin with Bayesian analysis that accounts for
variability in diagnostic accuracy and uncertainty in the transmission
dynamics.  As \citet{Wedderkopp2001} demonstrated, there is an
association between the proportion of \textit{S.}~Dublin
antibody-producing cows and the likelihood of the herd being
classified as seropositive.  The \textit{S.}~Dublin infection process
can be formulated as a dynamic state-space model (SSM)
\citep{Andrieu2010} where the hidden states represent the true
infection status of individual cows and BTM samples provide the
observable measurements.  Unlike hidden Markov models (HMM) that treat
infection status as a binary latent variable \citep{Madouasse2022,
  Roon2022, Meletis2023}, this formulation explicitly represents
latent infection and antibody production as a stochastic process.
This formulation would allow the use of Particle Markov-chain Monte
Carlo (PMCMC) methods \citep{Andrieu2010} for Bayesian,
simulation-based inference of the underlying infection dynamics.

From the infection dynamics, we can estimate the basic reproduction
number ($R_0$), a central concept in infectious disease epidemiology.
$R_0$ is defined as the average number of new cases of an infection
caused by one typical infected individual, in a population consisting
of susceptibles only \citep{Diekmann2010}.  If $R_0 < 1$, infection is
expected to die out.  However, because transmission is a stochastic
process there is no guarantee that infection will spread if $R_0 > 1$
\citep{Allen2008}.  Once infection has become established, depletion
of susceptibles (intrinsic factors) and implementation of
interventions (extrinsic factors) will affect the infection dynamics.
Then, the time-varying effective reproduction number ($R_t$) is a more
relevant metric, where $R_t$ is the actual average number of secondary
cases per primary case at time $t$ \citep{Nishiura2009}.  Thus, when
$R_t \ge 1$, the infection will typically remain endemic.

In addition to characterising regional differences in
\textit{S.}~Dublin prevalence, estimating $R_0$ and $R_t$ on Öland
could provide locally relevant data to support decision-making.  These
metrics depend not only on intrinsic factors but also on extrinsic
factors such as cattle demography, contact structure, country-specific
adoption of biosecurity practices \citep{Renault2021}, and differences
in the design of control programs across countries
\citep{Bergevoet2009, Nielsen2013b, Agren2015}.  Accounting for these
influences can help develop a more nuanced understanding of the
transmission dynamics of \textit{S.}~Dublin in Öland cattle.
Furthermore, previous studies have estimated $R_0$ in other regions,
such as Denmark \citep{Nielsen2007b} and the Netherlands
\citep{VanSchaik2007}, which highlights the importance of
region-specific estimates.  Thus, such analysis will provide valuable
insights into which interventions are most effective in limiting
spread, ultimately contributing to the development of effective
control and eradication strategies for \textit{S.}~Dublin in Swedish
cattle.

This study, therefore, aims to provide a more comprehensive
understanding of \textit{S.}~Dublin transmission dynamics on Öland by
estimating $R_0$ and $R_t$ using Bayesian simulation-based inference
of BTM sample results, and by evaluating the relative importance of
different transmission pathways to inform control strategies.

\section{Material and methods}
\label{material-and-methods}

\subsection{Livestock data}
\label{livestock-data}

The study incorporated all cattle events reported to the Swedish Board
of Agriculture from 1 April 2013 to 31 December 2024.  This dataset
included births, deaths, slaughter, imports, exports, sales, and
purchases of animals.  Each unique holding identifier in the reports
corresponds to a single geographical location where animals are kept,
which could represent a farm building, pasture, or other areas.

Cattle events were categorised into three types: enter, transfer, and
exit.  Enter events comprise births and imports, with new individuals
entering the model assumed to be susceptible. Transfer events
represent the movement of animals between holdings. Proportional
sampling was applied to the susceptible, infected, and
antibody-producing states to select individuals to be moved. Exit
events occur when animals leave a holding due to slaughter,
euthanasia, or export. Similarly, proportional sampling was applied to
the health states to select individuals to exit.  These categories
enabled us to incorporate population demographics and inter-herd
movements into the transmission model.

All holdings (n = \numprint{46026}) in the dataset were geopositioned,
except for seven holdings with missing coordinates, which were
excluded from the analysis.  To focus on Öland-specific data, events
occurring outside Öland were removed from the analysis.  All transfers
from other regions to Öland were reclassified as enter events of
susceptibles, and transfers from Öland to other regions were
reclassified as exit events.

\subsection{Bulk tank milk screenings}
\label{bulk-tank-milk-screenings}

BTM testing is a common diagnostic tool used to detect antibodies
against \textit{S.}~Dublin in dairy cattle \citep{Wedderkopp2001,
  Nielsen2012, Nyman2013}.  By analysing BTM samples, herd-level
seropositive status can be monitored non-invasively and
cost-effectively through BTM screening tests.  In total, results from
11 previously collected BTM screenings \citep{Agren2016, SVA2019,
  SVA2020, SVA2021, SVA2022, SVA2023, SVA2024} were used for the
Bayesian simulation-based inference. Only screenings from holdings
present in the Öland livestock dataset were included to ensure
consistency between the demographic data and the diagnostic
observations. The first screening (April 2013) was used to provide
input values for initialising the compartment model (within the SSM
framework) as described in \S\ref{model-posterior}, while the
remaining 10 screenings (Table~\ref{table:BTM-screenings}) were used
to measure agreement between the simulated outcome and observed
data. BTM samples were initially analysed using Prionics
PrioCHECK\textsuperscript{\textregistered} \textit{Salmonella} Ab
bovine ELISA (detecting O-antigens 1, 4, 5, 9, and 12) (Thermo Fisher
Scientific) to identify antibodies against any \textit{Salmonella}
serotype. Samples testing positive were subsequently analysed with
Prionics PrioCHECK\textsuperscript{\textregistered}
\textit{Salmonella} Ab bovine Dublin ELISA (detecting O-antigens 1, 9,
and 12) to determine the presence of antibodies specific to
\textit{S.}~Dublin. A corrected optic-density percent (ODC\%) of 20
had been used as a cut-off to classify a BTM sample as seropositive
\citep{Nyman2013}, which is lower than the 35 recommended by the
manufacturer. Information on the diagnostic sensitivity has not been
published, however, we assumed that it was consistent with what
\citet{Wedderkopp2001} reported. To account for uncertainty in test
performance, we conducted analyses under three ODC\% cut-off scenarios
as described in \S\ref{observation-process}. The BTM screenings were
collected on Öland between April 2013 to October 2024.

\begin{table}
  \centering
  \caption{BTM ELISA screening results of antibodies against
    \textit{S.}~Dublin in Swedish dairy herds on Öland. The 'Samples'
    column shows the number of BTM samples collected at each
    time-point, while 'Positive' indicates the number classified as
    seropositive. The posterior results (medians with 95\% credible
    intervals in parentheses) are from three ODC\% cut-off scenarios.}
  \label{table:BTM-screenings}
  \scriptsize
  \begin{tabular}{lccccc}
    \toprule
    & \multicolumn{2}{c}{Observed data} & \multicolumn{3}{c}{Posterior
      results by ODC\% cut-off scenario}\\
    \cmidrule(lr){2-3} \cmidrule(lr){4-6}
    & & & High Se, & Medium Se & Low Se,\\
    Date & Samples & Positive & low Sp & and Sp & high Sp\\
    \midrule
    2019-10-01 & 130 & 29 & 26 (20--33) & 27 (22--33) & 27 (17--32)\\
    2020-03-01 & 129 & 16 & 18 (13--24) & 18 (13--22) & 17 (13--22)\\
    2020-09-01 & 129 & 16 & 17 (12--22) & 16 (13--21) & 17 (13--21)\\
    2021-05-01 & 120 & 18 & 19 (14--24) & 19 (15--23) & 19 (15--23)\\
    2021-10-01 & 106 & 17 & 20 (15--26) & 20 (16--25) & 20 (16--25)\\
    2022-04-01 & 103 & 12 & 18 (13--23) & 18 (13--22) & 17 (13--22)\\
    2022-10-01 & 115 & 20 & 19 (14--24) & 19 (15--24) & 20 (15--24)\\
    2023-04-01 & 114 & 9 & 13 (9--22) & 12 (9--21) & 12 (8--20)\\
    2023-10-01 & 115 & 15 & 15 (10--20) & 15 (11--19) & 15 (11--18)\\
    2024-10-01 & 114 & 15 & 14 (6--23) & 14 (7--22) & 13 (6--21)\\
    \bottomrule
  \end{tabular}
\end{table}

\subsection{Compartment model}
\label{compartment-model}

\begin{figure}
  \centering
  \includegraphics[width=0.9\linewidth]{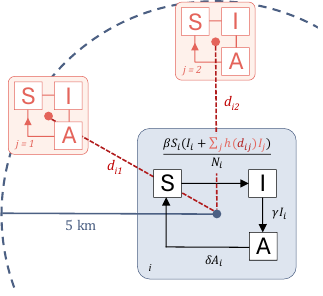}
  \caption{A schematic representation of the stochastic compartment
    model for \textit{S.}~Dublin in cattle. Animals within each
    holding transition between three health states: $S$ (susceptible),
    $I$ (infectious), and $A$ (antibody-producing). Between-holding
    transmission occurs via local spread, where infected animals at
    neighbouring holdings $j$, within a 5 km radius, contribute to the
    force of infection at holding $i$ through the transmission kernel
    $h(d_{ij})$, where $d_{ij}$ is the Euclidean distance between
    holdings $i$ and $j$.}
  \label{figure:model}
\end{figure}

Individuals were considered to belong to one of three health states:
$S$ susceptible; $I$ infectious; and $A$ antibody producing.  We
assumed that antibody production persists for a limited duration,
after which individuals revert to the susceptible state.  Furthermore,
it was assumed that nearby holdings influenced each other with
transmission of \textit{S.}~Dublin unrelated to livestock movements,
e.g.\ shared equipment, vehicles, and workers.  In addition, holdings
were connected by animal movements incorporated as transfer events
from real livestock data.  For holding $i$ and its neighbors $j$, the
transitions (Figure~\ref{figure:model}) between the states were
formulated as
\begin{align}
  \frac{dS_i}{dt} &= \delta A_i - \frac{\beta S_i ( I_i + \sum_j
    h(d_{ij}) I_j )}{N_i} \,, \label{eq:dS-dt} \\
  \frac{dI_i}{dt} &= \frac{\beta S_i ( I_i + \sum_j h(d_{ij}) I_j
    )}{N_i} - \gamma I_i \,, \label{eq:dI-dt} \\
  \frac{dA_i}{dt} &= \gamma I_i - \delta A_i \,, \label{eq:dA-dt}
\end{align}
with $N_{i} = S_{i} + I_{i} + A_{i}$, the total population size in
holding $i$.  Infected individuals became antibody producing at rate
$\gamma$, and the antibody producing capacity waned at rate $\delta$.
Furthermore, susceptible individuals became infected at a rate
proportional to the number of infected in holding $i$ and its
neighbors $j$, where $h(d_{ij})$ is the relative strength of
transmission to holding $i$ from holding $j$.  A transmission kernel
is commonly used to capture the distance-dependent probability of
disease spread between farms and has been applied to several different
livestock diseases \citep{Boender2007, Boender2023}.  In this study, a
Cauchy-type transmission kernel was adopted for the between-holding
transmission:
\begin{align}
  \label{eq:cauchy-transmission-kernel}
  h(d_{ij}) = \left[1+\left(d_{ij} / d_0 \right)^\alpha \right]^{-1} ,
\end{align}
where $d_{ij}$ was the Euclidean distance between holding $i$ and $j$,
$d_0$ is the distance where the relative strength of transmission is
half of its value at distance zero, and $\alpha$ the power parameter.
The value of $\alpha$ indicates whether the transmission is
short-ranged ($\alpha > 3$), intermediate-ranged ($2 < \alpha \le 3$),
or long-ranged ($\alpha \le 2$) \citep{Koeijer2011}.
\citet{Boender2023} estimated $\alpha$ in a truncated Lévy-walk kernel
(a related transmission kernel form with similar power-law decay
characteristics) for multiple livestock diseases where movement
restrictions had been applied as a control measure.  In the absence of
\textit{S.}~Dublin-specific estimates for local between-holding
transmission, we adopted the kernel shape estimated by
\citet{Boender2023} under movement restriction conditions.  This
approach isolates the local transmission component by excluding
movement-driven spread, which we model separately through event data.
The transmission kernel $h(d_{ij})$ was estimated for all $i \ne j$
such that $d_{ij} < 5$ km.  The 5 km distance cutoff was used because
presence of seropositive herds within that distance was found
associated with testing seropositive for \textit{S.}~Dublin in Sweden
\citep{Agren2016}.

The compartment model was implemented using the model parser function,
\texttt{mparse}, in the SimInf R package (version 10.1.0)
\citep{Widgren2019}. In addition, SimInf was extended to include
functionality for performing PMCMC inference informed by previous
works \citep{Bauer2016, Bronstein2023, Engblom2020}.

\begin{table*}
  \caption{Parameters and descriptions of the proposed compartment
    model for \textit{S.}~Dublin in cattle.}
  \label{table:model-parameters}
  \centering
  \footnotesize
  \begin{tabular}{llll}
    \toprule
    Variable  & Description & Value & Reference or prior $p(\theta)$\\

    \midrule

    $\gamma$ & Recovery rate per animal per day & $0.1$ &
    \citet{Xiao2005, VanSchaik2007}\\

    $\delta$ & Immunity-loss rate per animal per day & $1/90$ &
    \citet{VanSchaik2007}\\

    $\alpha$ & Power in the transmission kernel & $2.6$ &
    \citet{Boender2023}\\

    $d_{ij}$ & Distance between holding $i$ and $j$ (km) & Observed
    data & Geospatial data\\

    $\beta$ & Transmission rate per animal per day & Estimated &
    $\ln(\beta) \sim \mathcal{N}(\mu = -1.64, \sigma^2 = 1.74 \times
    10^{-1})$ \citep{Nielsen2007}\\

    $d_0$ & Half-distance in the transmission kernel (km) & Estimated
    & $\ln(d_0) \sim \mathcal{N}(\mu = -3.92 \times 10^{-2}, \sigma^2
    = 3.38 \times 10^{-1})$ \citep{Boender2023}\\

    \bottomrule
  \end{tabular}
\end{table*}

\subsection{Particle Markov-chain Monte
  Carlo methods}
\label{particle-markov-chain-monte-carlo}

An SSM is a Markov process where the internal state is hidden (also
referred to as latent) \citep{Andrieu2010}.  In this study, the hidden
state was the true number of $S$, $I$, and $A$ individuals.  Let
$\mathbf{x}_t = (S_t, I_t, A_t)$ denote the state variable of the
system at time $t$.  Moreover, due to the Markov property, the future
time evolution of the infectious process $\mathbf{x}_{t+1}$ is
predicted solely by the current state $\mathbf{x}_t$ and is
independent of previous states.  In addition, an SSM includes an
observation process to map the hidden state $\mathbf{x}_t$ to an
observation $\mathbf{y}_t$, a random variable depending on
$\mathbf{x}_t$.  This allowed us to relate the hidden state to
observed data comprising BTM samples from \textit{S.}~Dublin, which
enabled us to infer the underlying infection dynamics.  The SSM
combines the compartment model dynamics (Equations
\ref{eq:dS-dt}--\ref{eq:dA-dt}) with the observation process (BTM
screening data) to enable Bayesian inference of the hidden infection
states.  An SSM can be expressed in terms of probability density
functions as
\begin{align*}
  \begin{array}{ll}
    \boldsymbol{\theta} \sim \pi(\boldsymbol{\theta}) ,
    & \mathbf{x}_0 \sim \mu(\mathbf{x}_0 | \boldsymbol{\theta}) , \\
    \mathbf{x}_{t} \sim f(\mathbf{x}_{t} | \mathbf{x}_{t-1},
    \boldsymbol{\theta}) ,
    &\mathbf{y}_{t} \sim g(\mathbf{y}_{t} | \mathbf{x}_{t},
      \boldsymbol{\theta}),
  \end{array}
\end{align*}
where $\boldsymbol{\theta}$ denotes the parameters
(Table~\ref{table:model-parameters}) that govern the system's
evolution over time, and $\pi(\boldsymbol{\theta})$ is the prior
density of $\boldsymbol{\theta}$.  Furthermore, the initial state
$\mathbf{x}_0$ is distributed according to $\mu(\mathbf{x}_0 |
\boldsymbol{\theta})$, and $f(\mathbf{x}_{t+1} | \mathbf{x}_{t},
\boldsymbol{\theta})$ is a Markov kernel encoding the probability of
moving from state $\mathbf{x}_{t}$ at time $t$ to state
$\mathbf{x}_{t+1}$ at time $t+1$.  Similarly, $g(\mathbf{y}_{t} |
\mathbf{x}_{t}, \boldsymbol{\theta})$ denotes the probability density
of obtaining an observation $\mathbf{y}_{t}$, given that the current
state of the system is $\mathbf{x}_{t}$.

\begin{figure}
  \centering
  \includegraphics[width=\linewidth]{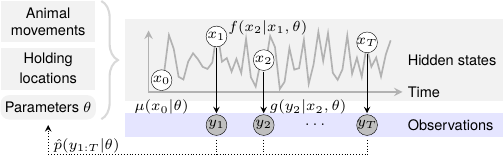}
  \caption{Schematic overview of the state-space model for
    \textit{S.}~Dublin in cattle. The initial state $\mathbf{x}_0$ is
    distributed according to $\mu(\mathbf{x}_0 |
    \boldsymbol{\theta})$. The hidden states $\mathbf{x}_t$ evolve
    over time according to the transition distribution $f(\mathbf{x}_t
    | \mathbf{x}_{t-1}, \boldsymbol{\theta})$, from which observed
    data are generated via the observation distribution
    $g(\mathbf{y}_t | \mathbf{x}_t, \boldsymbol{\theta})$. The
    posterior distribution for the model parameters
    $\boldsymbol{\theta}$ is estimated through sampling via the PMMH
    algorithm, using the BPF as an unbiased estimator for the
    likelihood.}
  \label{figure:state-space-model}
\end{figure}

Let $\mathbf{y}_{1:T} = (\mathbf{y}_1, \dots, \mathbf{y}_T)$ denote
the BTM screenings collected up to time $T$, and $\mathbf{x}_{0:T} =
(\mathbf{x}_0, \dots, \mathbf{x}_T)$ the hidden states.  Bayesian
inference, conditional on the observations $\mathbf{y}_{1:T}$, then
relies on the joint posterior distribution
\begin{align}
  p(\boldsymbol{\theta}, \mathbf{x}_{0:T} | \mathbf{y}_{1:T}) \propto
  \pi(\boldsymbol{\theta}) \mu(\mathbf{x}_0 | \boldsymbol{\theta})
  \prod_{t=1}^{T} f(\mathbf{x}_{t} | \mathbf{x}_{t-1},
  \boldsymbol{\theta}) \prod_{t=1}^{T} g(\mathbf{y}_{t} |
  \mathbf{x}_{t}, \boldsymbol{\theta}) .\nonumber
\end{align}
The equation is intractable for non-linear, non-Gaussian models,
making approximations and Monte Carlo methods necessary.  For SSMs,
PMCMC methods have been proposed as a useful approach for Bayesian
inference \citep{Andrieu2010}.  PMCMC consists of two parts.  Firstly,
a particle filter \citep{Gordon1993}, also referred to as sequential
Monte Carlo, is used to provide an unbiased approximation of the
marginal likelihood
$\hat{p}(\mathbf{y}_{1:T}\,|\,\boldsymbol{\theta})$, representing the
probability of observing the data given the chosen parameter value.
Secondly, using this estimate in a pseudo-marginal Metropolis-Hastings
(PMMH) algorithm, the Markov chain has the exact posterior
$p(\boldsymbol{\theta} | \mathbf{y}_{1:T})$ as the ergodic density
\citep{Andrieu2009}.  For a detailed explanation of how PMCMC works
and can be applied to disease dynamics, refer to \citet{Endo2019}.
Next, we provide a concise overview of the PMCMC workflow,
highlighting its key components and steps
(Figure~\ref{figure:state-space-model}).

A bootstrap particle filter (BPF) \citep{Gordon1993} was used to
sequentially simulate the hidden state in the holdings on Öland and
approximate the marginal likelihood
$\hat{p}(\mathbf{y}_{1:T}\,|\,\boldsymbol{\theta})$.  First, for a
given parameter value $\boldsymbol{\theta}$, N replicates of the SSM,
each initiated according to $\mu(\mathbf{x}_0 | \boldsymbol{\theta})$,
were constructed.  Each replicate of the SSM is denoted a particle and
represents a hypothetical value of the system state.  To evolve the
hidden state from $\mathbf{x}_0$ to $\mathbf{x}_1$,
i.e.,\ $f(\mathbf{x}_1 | \mathbf{x}_0, \boldsymbol{\theta})$, the SSM
for each particle was simulated forward to time $t_1$.  The likelihood
$g(\mathbf{y}_{1} | \mathbf{x}_{1}, \boldsymbol{\theta})$ to observe
$\mathbf{y}_1$ given $\mathbf{x}_1$ was then estimated for each
particle.  Moreover, to prevent a degenerate set of particles,
systematic resampling \citep{Li2015} was performed.  This was then
repeated sequentially for each time-step up to and including time $T$.
Ultimately, particles representing the posterior distribution and an
approximation of the marginal likelihood
$\hat{p}(\mathbf{y}_{1:T}\,|\,\boldsymbol{\theta})$ are obtained.

The BPF was then used as a sub-algorithm in PMMH:
\begin{enumerate}[1.]
\item Initiate the Markov chain by selecting an initial value for the
  parameters of interest, denoted as $\boldsymbol{\theta}^{(0)}$, at
  iteration $k = 0$. Next, run BPF with $N$ particles to estimate the
  approximate marginal likelihood,
  $\hat{p}(\mathbf{y}_{1:T}\,|\,\boldsymbol{\theta}^{(0)})$.  Finally,
  randomly select one trajectory, $\mathbf{x}_{1:T}^{(0)}$, from the
  BPF particles.
\item At iteration $k$, propose
  $\boldsymbol{\theta}^* \sim q(\boldsymbol{\theta}^k \,|\,
  \boldsymbol{\theta}^{k-1})$ from some proposal density
  $q(\cdot,\cdot)$.  An adaptive random walk \citep{Roberts2009} was
  used for the proposals of $\boldsymbol{\theta}$.
\item Run BPF with $\boldsymbol{\theta}^*$ to compute
  $\hat{p}(\mathbf{y}_{1:T} \,|\, \boldsymbol{\theta}^*)$ and randomly
  select one trajectory, $\mathbf{x}_{1:T}^*$.
\item With probability
  \begin{align}
    min \left[1, \frac{\hat{p}(\mathbf{y}_{1:T} \,|\,
        \boldsymbol{\theta^*})}{\hat{p}(\mathbf{y}_{1:T} \,|\,
        \boldsymbol{\theta^{(k-1)}})}
      \frac{q(\boldsymbol{\theta}^{(k-1)} \,|\,
        \boldsymbol{\theta}^*)}{q(\boldsymbol{\theta}^* \,|\,
        \boldsymbol{\theta}^{(k-1)})} \right] \nonumber
  \end{align}
  set $\boldsymbol{\theta}^{(k)} = \boldsymbol{\theta}^*$,
  $\mathbf{x}_{1:T}^{(k)} = \mathbf{x}_{1:T}^*$, and
  $\hat{p}(\mathbf{y}_{1:T} \,|\, \boldsymbol{\theta^{(k)}}) =
  \hat{p}(\mathbf{y}_{1:T} \,|\, \boldsymbol{\theta^*})$.  Otherwise,
  keep the values from the previous step.
\item Repeat steps 2--4 until the Markov-chain converges according to
  standard diagnostic criteria.
\end{enumerate}

The mixing of chains and the speed of convergence depend on the
variance of the approximated marginal likelihood
$Var(\hat{p}(\mathbf{y}_{1:T} \,|\, \boldsymbol{\theta}))$
\citep{Fearnhead2018}.  The variance relates to the number of
particles, and a variance of the log-likelihood estimate below 3 is
recommended \citep{Fearnhead2018}.  In our implementation, we used
$N=300$ particles to achieve a log-likelihood variance below this
threshold.

\subsection{Observation process}
\label{observation-process}

The BTM ELISA measures the ODC\% value from a serologically
representative pool, including all lactating cows on a dairy herd.
This ODC\% value is then classified as either seropositive or
seronegative against \textit{S.}~Dublin, depending on a certain
cut-off value used. Decreasing the ODC\% cut-off value increases the
diagnostic sensitivity (Se) but simultaneously decreases its
specificity (Sp).

To account for variability in diagnostic accuracy and explore the
impact of varying diagnostic performance on our understanding of the
underlying infection dynamics in cattle herds, three different ODC\%
cut-off scenarios were adopted: \textit{i}) Low Se, high Sp,
\textit{ii}) Medium Se and Sp, and \textit{iii}) High Se, low
Sp. Where the scenarios were assumed to correspond approximately to
ODC\% cut-off values of 40\%, 30\%, and 20\%, respectively. These
cut-off values correspond to different operating points on the
receiver operating characteristic (ROC) curve for the ELISA test, with
lower thresholds prioritising Se and higher thresholds prioritising
Sp.

To map the optic density (OD) value reported by \citet{Wedderkopp2001}
to ODC\% used to classify BTM samples in the screenings we proceeded
as follows. We start with a probabilistic model at the level of a
single animal; namely that one animal which sheds antibodies
contributes to the OD value according to $x \sim
\Gamma(\alpha_1,\theta)$, while one animal without antibodies
contributes $x' \sim \Gamma(\alpha_2,\theta)$, and where reasonably
$\alpha_1 \gg \alpha_2$ (but this is not assumed). In a mixture of $N$
animals where $PN$ animals shed antibodies and $(1-P)N$ do not, we
then get that the measured OD value is $y \sim
\Gamma(\alpha_1PN+\alpha_2(1-P)N, \theta/N)$. Given the data available
to us, $N$ is a latent variable and so has to be estimated.

In practice, it is reasonable to account for more sources of noise. To
allow for \emph{additive} noise in the form of a bias, we consider
that the recorded value $y$ is related to an ideal or ``true'' OD
value via the relation $y \sim OD+\nu$, where $\nu \sim
\Gamma(1/\sigma^2,\delta \sigma^2)$ satisfies $\Expect[\nu] = \delta$
and $\Var[\nu] = \sigma^2$. Reasonably, both parameters
$\{\delta,\sigma\}$ are expected to be small compared to the expected
signal $\approx \alpha_{1,2}\theta$. To further allow for
\emph{multiplicative} noise, we similarly consider data $y \sim OD
\times \varepsilon$, where $\varepsilon \sim \Gamma(1/\eta^2,\eta^2)$
satisfies $\Expect[\varepsilon] = 1$ and $\Var[\varepsilon] = \eta^2$,
again such that $\eta$ is a reasonably small relative error. These two
noise models are then straightforwardly combined, i.e., the recorded
data obeys $y_i \sim OD_i \times \varepsilon + \nu$, and assumed
i.i.d.~across the samples.

These additional noise terms require an approximation via moment
matching for the likelihood to be of closed form and is then valid for
relatively small noise $\{\delta,\sigma,\eta\}$. Conditioned on a
sample $i$ characterised by $\{P,N\}$ and given parameters
$\{\alpha_{1,2},\theta, \delta,\sigma,\eta\}$, we first compute
\begin{align}
  \label{eq:m1}
  \tilde{\alpha} &:= \alpha_1 P N+\alpha_2(1-P)N, \\
  \tilde{\theta} &:= \theta/N, \\
  \intertext{in terms of which}
    \theta^{*} &:=
  \frac{\tilde{\alpha} \tilde{\theta} ^2\left(1+\eta^2(1+\tilde{\alpha})\right)+\sigma^2}
                 {\tilde{\alpha} \tilde{\theta}+\delta}, \\
  \label{eq:m4}
  \alpha^{*} &:= \frac{\tilde{\alpha} \tilde{\theta} +\delta}{\theta^{*}},
\end{align}
resulting in a closed approximation in terms of the gamma density
$y_i \sim \Gamma(\alpha^{*},\theta^{*})$.

To estimate the parameters of the model, the BTM OD values and the
proportions of antibody-producing cows were first extracted from
Figure~1 in \citet{Wedderkopp2001} using Engauge Digitizer Version
12.1
(\url{https://markummitchell.github.io/engauge-digitizer/}). Since the
herd size $N$ was unknown to us it has to be treated as a latent
variable which leads us to the Expectation-Maximization (EM) algorithm
\citep{Dempster1977}. In our setting this becomes
\begin{description}
\item[Initialise] Form starting guesses for the model parameters
  $\{\alpha_{1,2},\theta, \delta,\sigma,\eta\}$ as well as a
  specification of the discrete latent states $N$.
\item[E-step] \emph{Given} parameters and latent states, compute
  likelihoods using \eqref{eq:m1}--\eqref{eq:m4} and the resulting
  gamma density. Normalise to form a discrete density over the latent
  states; these are the \emph{responsibilities} using
  EM-terminology. Use these to estimate the expected values for all
  latent variables.
\item[M-step] \emph{Given} the estimated latent variables, update the
  parameter estimates by numerically solving the corresponding maximum
  likelihood problem.
\item[Exit] Repeat the E- and M-steps until a tolerance criterion in
  the estimated parameters has been met.
\end{description}
In our case, \citet{Wedderkopp2001} have specified that $N \in
\{14,\ldots,141\}$. Additionally, the average herd size $= 55.5$ was
also given and to make use of this information, we implemented a
modification of the E-step due to \citet{Ganchev2007} which allows
constraints to be placed on the latent variables. Here, the density
over the latent variables is projected onto the known average size in
the sense of minimising the Kullback-Leibler divergence. In turn, the
projection problem can here be readily solved via Lagrange
multiplicators using the dual formulation, which is a scalar
zero-finding problem.

To compute the estimated likelihood $g(\mathbf{y}_{t} |
\mathbf{x}_{t}, \boldsymbol{\theta})$ of a BTM screening, the
probability that the OD value exceeds the cut-off is calculated for
each holding $i$ sampled in the BTM screening. The probability is
given by the cumulative distribution function
$F_{\textrm{OD}_{it}}(\textrm{cut-off} \mid \alpha^{*}_{it},
\theta^{*}_{it})$ of the gamma density:
\begin{align}
  \label{eq:sigmoid}
  P(\textrm{OD}_{it} > \textrm{cut-off}) = 1 -
  F_{\textrm{OD}_{it}}(\textrm{cut-off} \mid \alpha^*_{it},\theta^*_{it}) .
\end{align}
Because each BTM sample has its own success probability, all samples
in a BTM screening follow a Poisson binomial distribution as the sum
of independent Bernoulli trials. The R package \texttt{poibin}
\citep{Hong2024} was used to compute this combined likelihood. The
characteristic sigmoid shape of Equation~\eqref{eq:sigmoid} is shown
in Figure~\ref{figure:probability-bulk-milk-sample} for the three
ODC\% cut-off scenarios and a few representative herd sizes.

\subsection{Model posterior}
\label{model-posterior}

The parameters to be estimated in the model were
$\boldsymbol{\theta}=\{\beta,d_0\}$.  Initial values of $d_0$ were
chosen based on previously calculated $d_0$ for a range of outbreaks
after imposing movement bans \citep{Boender2023}, i.e.,\ the same data
as was used to form the prior of $d_0$. Furthermore, after initial
exploration of model simulations and parameterisation, plausible
corresponding initial values of $\beta$ were chosen for each $d_0$.
In total there were four starting points, to enable four separate
chains of PMCMC.  Because both parameters must be positive, they were
transformed to the log-scale during the parameterisation process.  For
one of the starting points, 500 estimates of the log-likelihood were
generated with $N=300$ particles in the BPF.  The variance
$Var(\hat{p}(y_{1:T}|\theta))$ of these estimates was then calculated.

For the half-distance $d_0$, a lognormal distribution was assumed
where the mean and standard deviation of the distribution on the log
scale were based on the values reported for
$d_0 = \{0.64, 1.33, 0.76, 2.7, 0.78, 0.58\}$ by \citet{Boender2023}
after movement restrictions had been applied.  For the transmission
rate $\beta$, a lognormal distribution of the prior was also assumed.
\citet{Nielsen2007} reported $R_0 = \{1.1, 2.6, 1.8, 2.7\}$.  Those
values were multiplied by $\gamma$ to obtain an estimate of $\beta$,
and the mean and standard deviation of the distribution on the log
scale were calculated.

The initial state $\mu(\mathbf{x}_0 | \boldsymbol{\theta})$ at
$t_0=\textrm{2013-04-01}$ was set according to the BTM screening
prevalence in Öland at that time \citep{Agren2016}.  In holdings with
$N_i(t_0)>0$, 15\% were randomly sampled to contain infected
individuals.  The state in initially infected holdings was set as
follows: $S_i(t_0)=0.95N_i(t_0)$, which represents 95\% of the cattle
population being susceptible to infection; and
$I_i(t_0)=0.05N_i(t_0)$, representing 5\% of the cattle being
initially infected with \textit{S.}~Dublin.  The initial state in the
other non-infected holdings was $S_i(t_0)=N_i(t_0)$, $I_i(t_0)=0$, and
$A_i(t_0)=0$.  After initialisation, the simulation ran for 2374 days
before the next BTM screening in October 2019.

Four chains of PMCMC were run for \numprint{10000} iterations each
(totalling \numprint{40000} iterations).  The first \numprint{5000}
iterations of each chain were discarded as burn-in, leaving
\numprint{5000} samples per chain.  This resulted in a total of $N =$
\numprint{20000} posterior samples.  The PMCMC inference was performed
independently for each of the three ODC\% cut-off scenarios (High Se,
low Sp; Medium Se and Sp; Low Se, high Sp), resulting in three
distinct posterior distributions for $\beta$ and $d_0$.  To assess
convergence, the effective sample size (ESS) metrics bulk-ESS and
tail-ESS, and the alternative rank-based $\hat{R}$, were calculated as
suggested by \citet{Vehtari2021}.  The R package \texttt{rstan}
\citep{rstan} was used to calculate these metrics.  It is recommended
that $\hat{R} < 1.01$ and that both bulk-ESS and tail-ESS are at least
100 per Markov chain for a reliable posterior \citep{Vehtari2021}.

\begin{figure}
  \centering
  \includegraphics[width=\linewidth]{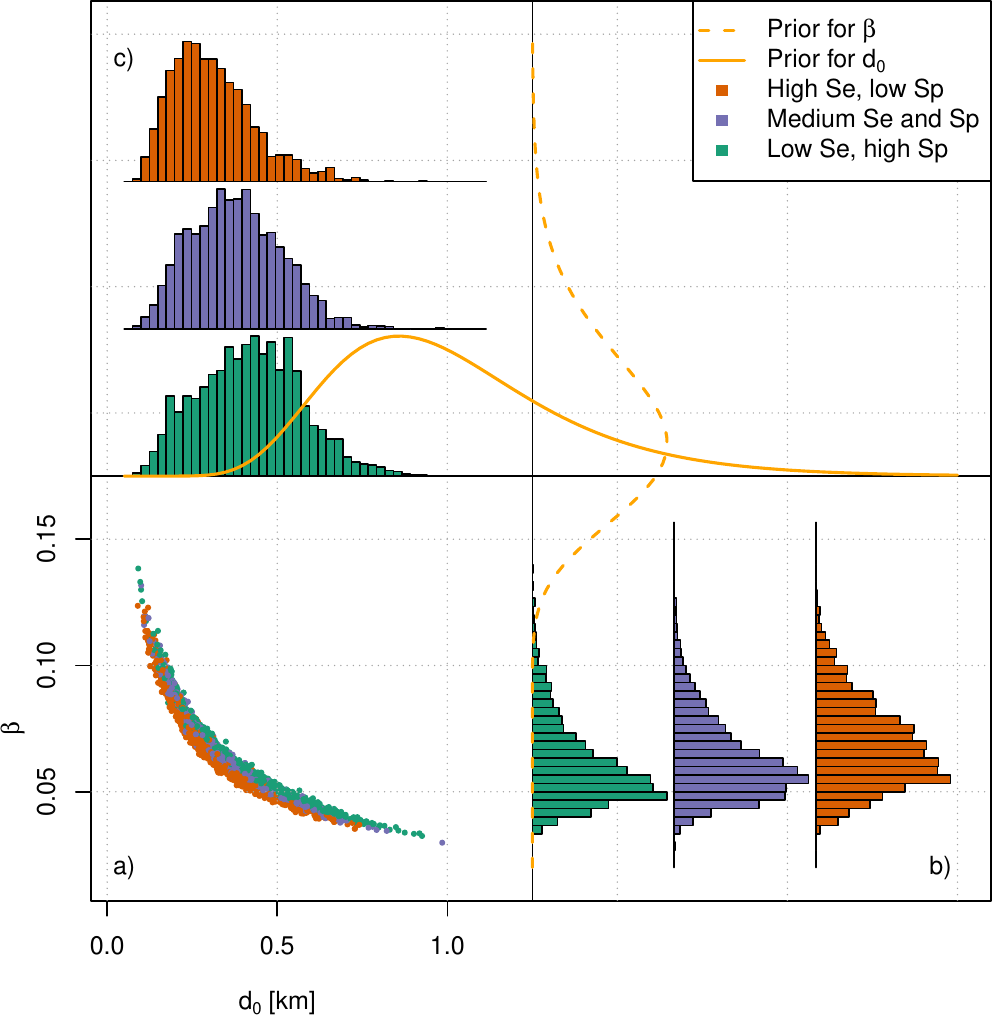}
  \caption{Approximate posterior distribution of the compartment model
    parameters for \textit{S.}~Dublin in cattle on Öland, estimated
    via particle Markov-chain Monte Carlo. Results are shown for three
    ODC\% cut-off scenarios. a) Joint posterior distribution (1000
    samples per scenario). b, c) Marginal posterior distributions for
    the transmission rate ($\beta$) and half-distance ($d_0$) in the
    transmission kernel, respectively. Orange lines indicate the prior
    distributions.}
  \label{figure:posterior}
\end{figure}

\subsection{Synthetic Data Generation for Validation}
\label{synthetic-data-generation-for-validation}

To validate the model and compare simulations with observed data, we
generated synthetic observations from the final posterior samples
obtained after the PMCMC chains converged. For each of the
\numprint{20000} posterior samples (\numprint{5000} per chain across
four chains), and for each of the three ODC\% cut-off scenarios, we
calculated the probability $p_i$ that a holding $i$ would test
positive at the time of a BTM screening. This probability was derived
from the proportion of antibody-producing cows ($A_i/N_i$) in the
simulated state and the diagnostic characteristics of the specific
scenario. We then drew a uniform random number $U \sim
\text{Uniform}(0,1)$ for each holding. If $U < p_i$, the holding was
classified as seropositive; otherwise, it was classified as
seronegative. The mean and 95\% credible intervals of these synthetic
classifications across all posterior samples were calculated and
presented in Table~\ref{table:BTM-screenings} alongside the observed
data. This step facilitated a direct visual and statistical comparison
between the model's posterior predictions and the real-world screening
results, confirming the model's ability to reproduce observed
prevalence patterns.

\subsection{Basic reproduction number}
\label{basic-reproduction-number}

To estimate the basic reproduction number $R_0$, numerical simulations
were employed using posterior parameter samples from the PMCMC
inference. The transition equations \eqref{eq:dS-dt}--\eqref{eq:dA-dt}
were adapted to enable counting the number of secondary cases
generated by introducing one infected individual into a
holding. Specifically, the model was modified to include a state for
the introduction of an infected individual ($I_{0i}$) into a
previously susceptible population:
\begin{align}
  \frac{dS_i}{dt} &= \delta A_i - \frac{\beta S_i [I_i + I_{0i} +
      \sum_j h(d_{ij}) (I_j + I_{0j})]}{N_{0i}} ,\nonumber\\
  \frac{dI_i}{dt} &= \frac{\beta S_i [I_i + I_{0i} + \sum_j h(d_{ij})
      (I_j + I_{0j})]}{N_{0i}} - \gamma I_i ,\nonumber\\
  \frac{dI_{0i}}{dt} &= - \gamma I_{0i} ,\nonumber\\
  \frac{dA_i}{dt} &= \gamma (I_i + I_{0i}) - \delta A_i ,\nonumber
\end{align}
with $N_i = S_i + I_i + I_{0i} + A_i$, the total population size in
holding $i$. With this reformulation, it was possible in SimInf to
count the number of individuals infected by the introduced case. For
each holding $i$, \numprint{1000} posterior samples were randomly
selected from each of the three ODC\% cut-off scenarios, resulting in
\numprint{3000} total samples per holding. Then, for each sample, a
time-point between 1 January 2020 to 31 December 2024 was randomly
selected, where the population size $N_i > 0$. At this time-point, in
a fully susceptible population, one susceptible individual ($S$) was
moved to become infected ($I_{0i}$), and the simulation continued
until 31 December 2024. The number of individuals infected by the
introduced case was recorded for each holding $i$ and posterior
sample. To obtain a distribution of $R_0$, representing the expected
number of secondary cases generated by introducing one infected
individual into a holding $i$ in Öland, the estimates per holding were
averaged ($\overline{R_0^i}$) to generate a summary statistic.

\subsection{Effective reproduction number}
\label{effective-reproduction-number}

To investigate how changes in the model affect the effective
reproduction number $R_t$, three different scenarios were simulated to
compare their effects on $R_t$. Each scenario was further divided into
the three ODC\% cut-off sub-scenarios. The baseline scenario
represents the estimated $R_t$ from posterior samples. A second
scenario explored the impact of restricting the movements of infected
cattle, while a third scenario examined the effect of decreasing the
half-distance $d_0$ in the between-holding transmission kernel.

For each posterior sample, 50 replicates were simulated to ensure
robust estimation of the mean trajectory and to minimise the influence
of process variability. Daily incidence ($i_t$) and prevalence
($I_{t-1}$) were averaged across the 50 replicates for each posterior
sample. The instantaneous effective reproduction number, $R_t$, was
then calculated for each of the smoothed trajectories based on the
simulation output for the period from 1 January 2020 to 31 December
2024. Since the simulator records the exact number of new infections
and the exact number of infected animals in the previous time step,
$R_t$ can be derived directly from the model dynamics as $R_t=
\bar{i}_{t} / (\gamma \bar{I}_{t-1})$ where $\bar{i}_{t}$ and
$\bar{I}_{t-1}$ denote the averaged daily incidence and the number of
infected animals the previous day, respectively, and $\gamma =
0.1~\text{day}^{-1}$ is the recovery rate (implying a mean infectious
period of 10 days). This formulation represents the average number of
secondary cases generated per infected individual per day, scaled by
the recovery rate. Finally, the median and 95\% simulation envelope
(2.5th to 97.5th percentiles) were computed across the posterior
sample-level estimates. This two-stage aggregation approach separates
the variability due to parameter uncertainty from the fine-scale
process noise inherent in individual transmission events.

First, the baseline $R_t$ was estimated by simulating 50 replicates
for each of the \numprint{1000} random posterior samples from each
ODC\% cut-off scenario.

Second, the impact of restricting the movements of infected cattle was
explored. In SimInf, scheduled transfer events (animal movements) are
processed when simulation time reaches any of these events. This
process involves sampling individuals from compartments in a source
holding and moving them to the same compartments in a destination
holding. The probability of sampling an individual from each
compartment is
\begin{align}
  \label{eq:movement-probability}
  p_S &= w_S S / (w_S S + w_I I + w_A A) ,\nonumber\\
  p_I &= w_I I / (w_S S + w_I I + w_A A) ,\nonumber\\
  p_A &= w_A A / (w_S S + w_I I + w_A A) ,\nonumber
\end{align}
where $w_S$, $w_I$, and $w_A$ are the probability weights. For the
baseline $R_t$ estimation, equal probability weights ($w_S=w_I=w_A=1$)
were used, which meant that the probability of moving an individual
from a compartment was proportional to the number of individuals in
that compartment. To restrict movements of infected cattle, the
probability weight for infected animals was reduced for movements
occurring after 1 January 2020 ($w_S=w_A=1$; $w_I=0.01$). Then, the
effective reproduction number after applying movement restrictions,
denoted $R_t^m$, was estimated by simulating 50 replicates for each of
the \numprint{1000} random posterior samples from each ODC\% cut-off
scenario.

Third, the effect of decreasing the half-distance $d_0$ by 50\% in the
between-holding transmission kernel after 1 January 2020 was
explored. The effective reproduction number, denoted $R_t^d$, was
estimated by simulating 50 replicates for each of the \numprint{1000}
random posterior samples from each ODC\% cut-off scenario.

\subsection{Posterior predictive simulations}
\label{posterior-predictive-simulations}

To quantify the relative contribution of transmission routes for
\textit{S.}~Dublin, we calculated the proportion of daily incident
cases attributable to local spread (external transmission) versus
within-holding transmission (internal). For each of the
\numprint{1000} random posterior samples from each of the three ODC\%
cut-off scenarios, we simulated one trajectory covering the period
from 1 January 2020 to 31 December 2024. The proportion was computed
as the ratio of new infections generated by local spread to the total
number of new infections on each day. In time steps with no incident
cases, this proportion was defined as 0\%. The median and 95\% central
credible interval (CrI) were estimated for this proportion across the
\numprint{1000} simulated trajectories for each scenario.

Similarly, to assess the spread of \textit{S.}~Dublin within the
population, we calculated three prevalence measures for each of the
three ODC\% scenarios for the same period: (1) holding-level
prevalence (proportion of infected holdings), (2) within-holding
prevalence (proportion of infected ($I$) and antibody-producing ($A$)
animals conditioned on the presence of at least one infected
individual in the holding), and (3) individual-level prevalence
(proportion of infected ($I$) and antibody-producing ($A$) animals
across all holdings). All measures were summarised using the median
and 95\% CrI across the \numprint{1000} simulated trajectories per
scenario.

\begin{figure}
  \centering
  \includegraphics[width=\linewidth]{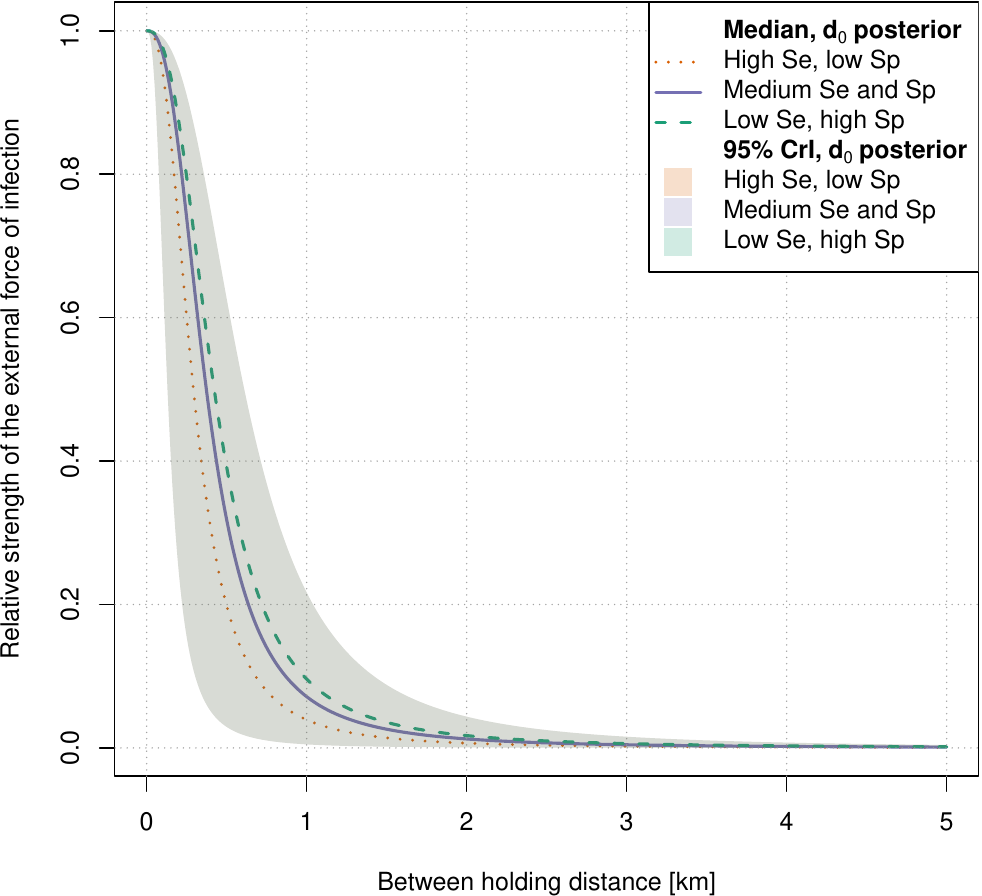}
  \caption{Distribution of the estimated transmission kernel for
    \textit{S.}~Dublin in cattle on Öland, Sweden. The transmission
    kernel represents the relative strength of the external force of
    infection from an infected neighbour as a function of
    between-holding distance. The half-distance ($d_0$) was estimated
    via particle Markov-chain Monte Carlo. Results are shown for three
    ODC\% cut-off scenarios. Lines represent the median estimates for
    each scenario, while grey areas denote the 95\% credible
    intervals, which overlap substantially between scenarios.}
  \label{figure:cauchy-kernel}
\end{figure}

\begin{figure}[!t]
  \centering
  \includegraphics[width=\linewidth]{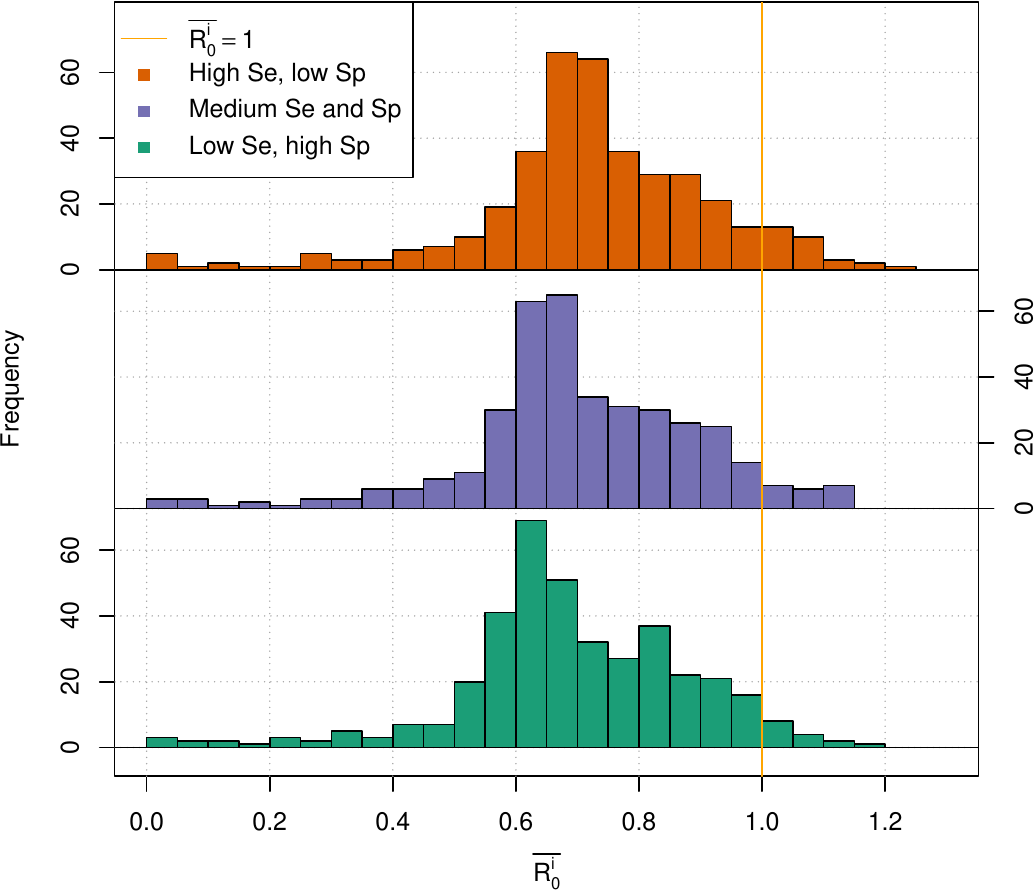}
  \caption{Histogram of the average basic reproduction number
    $\overline{R_0^i}$ for \textit{S.}~Dublin in cattle across
    holdings on Öland, Sweden. Parameters were estimated via particle
    Markov-chain Monte Carlo. Results are shown for three ODC\%
    cut-off scenarios. The orange line indicates the threshold value
    $\overline{R_0^i}=1$.}
  \label{figure:R_0}
\end{figure}

\section{Results}
\label{results}

\subsection{Livestock data}
\label{results-livestock-data}

In total, the cleaned events data for Öland comprised \numprint{465}
holdings, \numprint{276406} exit events, \numprint{274103} enter
events, and \numprint{106150} transfer events. On average, 28.5
(median: 19; IQR: 8--37) animals were moved per day within Öland
during the study period. Additionally, the average holding size was
189 animals (median: 110; IQR: 29--285). For the holdings included in
the BTM screenings, the average total herd size was 319 animals
(median: 285; IQR: 136--456).

\subsection{Observation process}
\label{results-observation-process}

\begin{figure}
  \centering
  \includegraphics{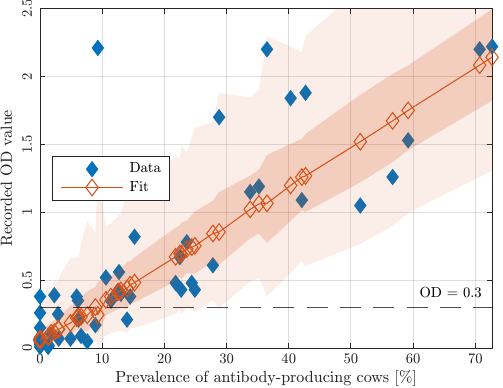}
  \caption{Probabilistic model of OD value as a function of the
    prevalence of antibody-producing cows. Scattered data points are
    from \citet{Wedderkopp2001}, who proposed an OD cut-off value of
    0.3. Our model is formulated using gamma distributions and
    includes an explicit dependence on herd size. The fitted model is
    shown as expected values with 50\% and 95\% credible intervals.}
\label{fig:GamODfit}
\end{figure}

In practice, the procedure described in \S\ref{observation-process}
converges well and produces reasonable fits to the data
(Figure~\ref{fig:GamODfit}). The final model has the estimated
parameters:
\begin{align}
  \alpha_1 &= \num{3.76e-1}, \quad \alpha_2 = \num{5.72e-03}, \quad
  \theta = 7.77, \\
  \delta &= \num{4.18e-02}, \quad \sigma = \num{2.08e-09}, \quad \eta
  = \num{2.82e-05}. \nonumber
\end{align}
Notably, $\alpha_2\theta \approx 1.1 \times \delta$, i.e., the
expected background shedding of one healthy individual is about the
same magnitude as the estimated bias $\delta$. Importantly, the signal
quotient $\alpha_1/\alpha_2 \approx 65.7$, indicating that
antibody-producing individuals are indeed associated with a clear
signal.

\begin{figure}
  \centering
  \includegraphics{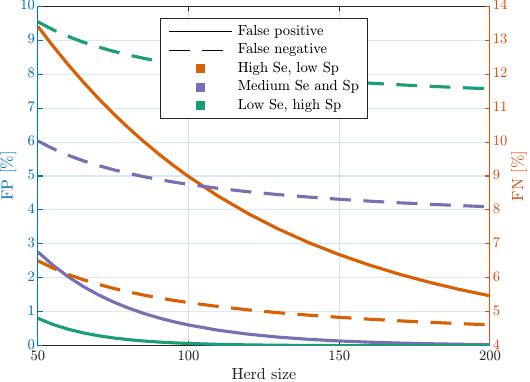}
  \caption{False positive/negative trade-off of the proposed OD value
    model under different sensitivity scenarios as a function of herd
    size. In low-prevalence scenarios and for small herd sizes, the
    relatively high false positive rate in the High Se scenario
    complicates model calibration, whereas the Low Se scenario may
    miss true positives.}
\label{fig:GamFPFN}
\end{figure}

\begin{figure}[!t]
  \centering
  \includegraphics[width=\linewidth]{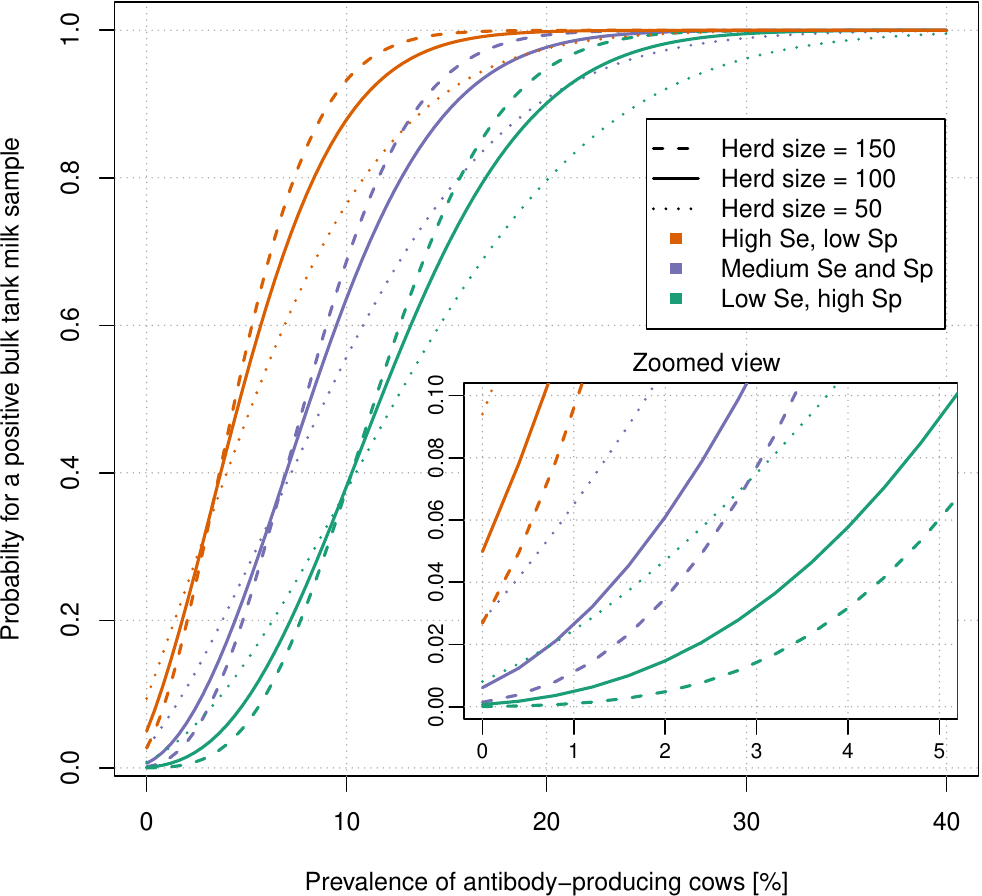}
  \caption{Classification of bulk tank milk (BTM) samples as
    seropositive against \textit{S.}~Dublin: estimated cumulative
    distribution function (CDF). The CDF shows the probability of
    seropositivity based on optical density (OD), with separate curves
    for different herd sizes and OD cut-off values. This probability
    depends on the OD cut-off value, herd size, and the proportion of
    \textit{S.}~Dublin antibody-producing cows in the dairy herd. The
    CDF was estimated using data from \citet{Wedderkopp2001},
    independently of the BTM screening data.}
\label{figure:probability-bulk-milk-sample}
\end{figure}

Figure~\ref{fig:GamODfit} compares the model predictions with the
original data from \citet{Wedderkopp2001}, while
Figure~\ref{fig:GamFPFN} illustrates the false positive and false
negative rates as functions of herd size. Across the ten screening
dates, the model demonstrates strong calibration: nine out of ten
observed prevalence values fall within the 95\% credible intervals for
all three ODC\% cut-off scenarios, see
Table~\ref{table:BTM-screenings}. The single exception (2022-04-01),
where the observation fell below the lower bound of the credible
intervals in all scenarios. Given the average herd size on Öland
established earlier, these results indicate that the model reliably
captures the underlying prevalence dynamics across the full range of
sensitivity assumptions.

\subsection{Model posterior}
\label{results-model-posterior}

Summary of the posterior distributions of $\theta=\{\beta,d_0\}$ are
shown in Table~\ref{tab:posterior}, with the marginal and joint
distributions visualised in
Figure~\ref{figure:posterior}. Figure~\ref{figure:cauchy-kernel} shows
the relative strength of transmission according to
Equation~\ref{eq:cauchy-transmission-kernel} for distances up to 5 km,
based on the mean and 95\% CI of the posterior of $d_0$, for the three
ODC\% cut-off scenarios. The estimated variance of the log-likelihood
approximations $Var(\hat{p}(\mathbf{y}_{1:T} \,|\,
\boldsymbol{\theta}))$ was 0.3 (Low Se, high Sp), 0.4 (Medium Se and
Sp), and 0.2 (High Se, low Sp), evaluated at the posterior median
values of $\boldsymbol{\theta}$. All posterior chains converged with
$\hat{R} < 1.01$ and bulk-ESS/tail-ESS values exceeding 100 per chain,
as shown in Table~\ref{tab:posterior}.

\begin{table}
  \caption{Marginal posterior distribution of the compartment model
    parameters for \textit{S.}~Dublin in cattle on Öland, Sweden,
    estimated via particle Markov-chain Monte Carlo. Results are shown
    for three ODC\% cut-off scenarios. The table includes the
    effective sample size (ESS) metrics bulk-ESS and tail-ESS, and the
    rank-based $\hat{R}$ to assess convergence.}
  \label{tab:posterior}
  \centering
  \scriptsize
  \begin{tabular}{llcccccc}
    \toprule
    & & & & & \multicolumn{2}{c}{ESS$^\dagger$}\\
    \cmidrule(lr){6-7}
    Parameter & ODC\% scenario & Median & 95\% CrI & $\hat{R}^\dagger$ &
    bulk & tail\\
    \midrule
    $\beta$ & High Se, low Sp & 0.066 & 0.041--0.108 & 1.001 & \numprint{1722} & \numprint{2248} \\
    $\beta$ & Med Se and Sp & 0.059 & 0.041--0.102 & 1.004 & \numprint{1329}  & \numprint{1608} \\
    $\beta$ & Low Se, high Sp & 0.057 & 0.039--0.101 & 1.002 & \numprint{1425} & \numprint{1692} \\
    $d_0$ [km] & High Se, low Sp & 0.292 & 0.130--0.610 & 1.002 & \numprint{1650} & \numprint{1983} \\
    $d_0$ [km] & Med Se and Sp & 0.373 & 0.155--0.652 & 1.004 & \numprint{1291} & \numprint{1651} \\
    $d_0$ [km] & Low Se, high Sp & 0.422 & 0.157--0.733 & 1.002 & \numprint{1401} & \numprint{1523} \\
    \bottomrule
  \end{tabular}
  \begin{minipage}{\textwidth}
    $^\dagger$ $\hat{R}$ and ESS metrics were assessed based on
    $\log(\beta)$ and $\log(d_0)$.
  \end{minipage}
\end{table}

\subsection{Basic reproduction number}
\label{results-basic-reproduction-number}

The distributions of simulated values of $\overline{R_0^i}$ (per
holding) are illustrated in Figure~\ref{figure:R_0}. In total, there
were 386 active holdings during the simulated period between 1 January
2020 to 31 December 2024. The mean values of $\overline{R_0^i}$ were
0.73 for High Se, low Sp, 0.71 for Medium Se and Sp, and 0.69 for Low
Se, high Sp. Some holdings had $\overline{R_0^i} > 1$: specifically,
29 holdings for High Se, low Sp, 20 for Medium Se and Sp, and 15 for
Low Se, high Sp.

\subsection{Effective reproduction number}
\label{results-effective-reproduction-number}

\begin{figure}
  \centering
  \includegraphics[width=\linewidth]{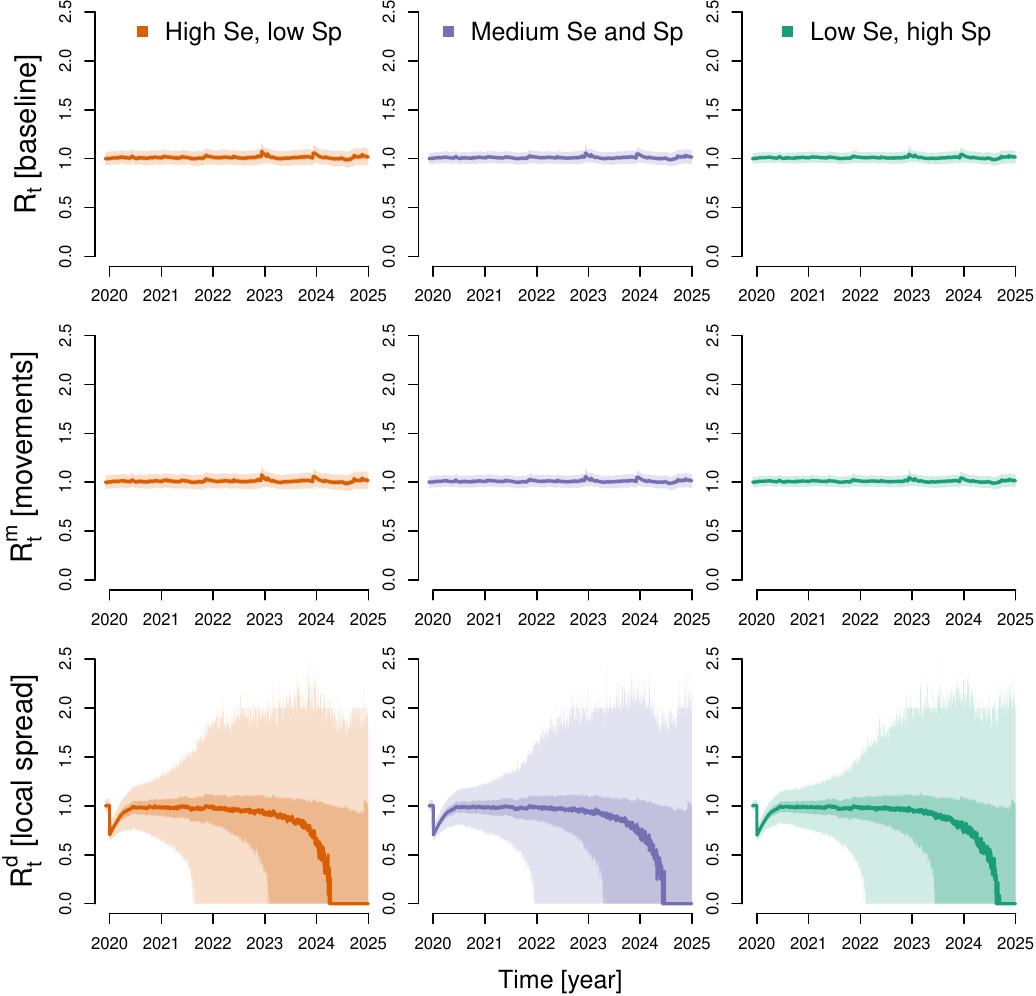}
  \caption{Simulated effective reproduction number ($R_t$) for
    \textit{S.}~Dublin in cattle on Öland, Sweden, across a matrix of
    screening thresholds and intervention strategies. Columns
    represent three ODC\% cut-off scenarios, while rows represent the
    intervention scenarios: (Top) Baseline, (Middle) Movement
    restrictions of infected cattle, and (Bottom) Reduced local
    transmission.  Solid lines indicate the median $R_t$ across
    \numprint{1000} posterior predictive trajectories (each derived
    from the average of 50 stochastic replicates per posterior
    sample).  Darker and lighter shaded regions denote the 50\% and
    95\% simulation envelopes, respectively.}
\label{figure:R-effective}
\end{figure}

The simulated effective reproduction numbers ($R_t$, $R_t^m$, and
$R_t^d$) are illustrated in Figure~\ref{figure:R-effective}. For the
baseline scenario ($R_t$) and the movement restriction scenario
($R_t^m$), the median effective reproduction number remained close to
1 across all three ODC\% cut-off scenarios (High Se, low Sp; Medium Se
and Sp; Low Se, high Sp), exhibiting no clear temporal trend.

In contrast, the scenario with a 50\% reduction in the transmission
half-distance ($R_t^d$) displayed a distinct divergence between the
central tendency and the uncertainty bounds. While the median declined
toward zero---indicating a successful suppression of the average
transmission potential---the 95\% simulation envelope widened
considerably, reaching values as high as 2.5. This expansion reflects
increased stochastic variability driven by the larger pool of
susceptible animals remaining in the population. Under local
transmission reduction, \textit{S.}~Dublin is suppressed on average,
but the persistence of high susceptibility allows for sporadic,
localised outbreaks in specific replicates, creating a ``highly
variable'' distribution of $R_t^d$ values. Consequently, while the
intervention effectively lowers the expected transmission, it
coincides with a broader range of possible outcomes, including
occasional high-transmission events.

Across the three ODC\% cut-off scenarios, a temporal gradient was
observed in the timing of median $R_t^d$ decline. The High Se, low Sp
scenario reached median $R_t^d \approx 0$ approximately 6 months
earlier than the Low Se, high Sp scenario, with the Medium Se and Sp
scenario falling between these extremes.

\subsection{Posterior predictive simulations}
\label{results-posterior-predictive-simulations}

The proportion of daily incident cases attributable to local spread
was estimated to be approximately 50--60\% across all diagnostic
scenarios. Specifically, the median proportion and 95\% CrI was 59\%
(30--83) for the Low Se, high Sp scenario, 56\% (27--85) for Medium Se
and Sp, and 53\% (0--89) for the High Se, low Sp scenario. Despite
differences in the underlying parameter estimates across scenarios,
the relative contribution of local spread remained stable. This
indicates that local transmission accounts for roughly half of all new
infections, and this finding is robust to uncertainty in the
diagnostic sensitivity and specificity assumptions.

However, the estimated holding-level prevalence (the proportion of all
herds harbouring at least one infected animal) differed across
diagnostic scenarios: 15\% (5--26) for Low Se, high Sp, 12\% (3--22)
for Medium Se and Sp, and 8\% (2--18) for High Se, low Sp.  In
contrast, among holdings with at least one infected animal, the
within-holding prevalence remained relatively stable across scenarios.
Specifically, the median proportion of infected animals ($I$) within
these infected holdings was 2\% (0--25) for Low Se, high Sp, 2\%
(0--23) for Medium Se and Sp, and 3\% (0--21) for High Se, low
Sp. Furthermore, the median proportion of antibody-producing animals
($A$) within these same infected holdings was notably higher: 17\%
(0--74) for Low Se, high Sp, 18\% (0--75) for Medium Se and Sp, and
21\% (0--77) for High Se, low Sp. In summary, while test sensitivity
had an impact on the estimated number of infected holdings, the
intensity of infection within those infected holdings was consistent
across scenarios.

Finally, the median prevalence of infected animals on Öland was 0.5\%
(0--1) for Low Se, high Sp, 0.4\% (0--1) for Medium Se and Sp, and
0.3\% (0--1) for High Se, low Sp, while the median prevalence of
antibody-producing animals was 4\% (2--7) for Low Se, high Sp, 3\%
(1--7) for Medium Se and Sp, and 3\% (1--6) for High Se, low Sp.

\section{Discussion}
\label{discussion}

This study applied Bayesian simulation-based inference to estimate two
key epidemiological quantities---the basic reproduction number $R_{0}$
and the time-varying effective reproduction number $R_{t}$---for
\textit{S.}~Dublin transmission on Öland.  In contrast, many studies
use Bayesian HMMs that predict only herd-level infection status,
treating the true infection state as a latent binary variable inferred
from imperfect diagnostic tests \citep{Madouasse2022, Roon2022,
  Meletis2023}.  Although analytically convenient, such HMMs assume a
simplistic underlying process and cannot capture the complex dynamics
of pathogen spread within and between holdings.  To overcome this
limitation, we formulate an SSM that explicitly represents latent
infection and antibody production as a stochastic process, linking
these latent states to observed test results via a measurement model.
Because the resulting likelihood is intractable, we employ PMCMC to
obtain unbiased posterior estimates of both model parameters and
latent trajectories.  Thus, our approach refines inference beyond HMMs
that use a binary latent infection state, providing a more nuanced
understanding of \textit{S.}~Dublin transmission dynamics in cattle
populations.

Our posterior predictive estimates of the reproduction numbers further
elucidate the transmission dynamics of the pathogen.  Specifically,
the effective reproduction number was estimated at $R_t \approx 1$.
An $R_t$ value close to unity implies that transmission is neither
expanding nor contracting \citep{Nishiura2009}, consistent with a
stable, endemic presence of \textit{S.}~Dublin in the Öland cattle
population under the current management and biosecurity practices.
However, this population-level stability masks significant
heterogeneity at the herd level.  While most holdings in the study
have an $R_0 < 1$ (suggesting the infection would die out in
isolation), a subset of holdings exhibit $R_0 > 1$.  This indicates
that these specific holdings act as potential reservoirs capable of
sustaining transmission independently, posing a persistent risk of
spread regardless of the current endemic status.  These high-risk
holdings likely drive the overall persistence of the pathogen in the
region.

Comparing our results with previously published $R_{0}$ estimates, we
find that our values are lower than those reported by
\citet{Xiao2005}, \citet{Nielsen2007}, and \citet{VanSchaik2007}.
This discrepancy may stem partly from methodological
differences---e.g., \citet{Nielsen2007} estimated $R_{0}$ in young
dairy calves---or from country-specific variations in the adoption of
biosecurity practices among farmers \citep{Renault2021}.  For
instance, Sweden has a long tradition of preventive control of
Salmonella; legislated national Salmonella-control measures have been
in place since the 1960s \citep{Agren2016}.

In the compartment model, \textit{S.}~Dublin transmission occurs via
three distinct mechanisms: internal spread within a holding, local
spread between proximate holdings, and transmission via animal
movements. While animal movements are explicitly modeled, they do not
generate new infections de novo; rather, they function solely as a
transport mechanism that moves already-infected animals between
holdings. This introduces the pathogen to a new location or augments
the existing infectious pool, irrespective of whether the recipient
herd was previously uninfected or already harboring the
infection. Posterior predictive estimates indicate that internal and
local spread contribute almost equally to the overall force of
infection, each accounting for roughly 50\% of daily new
infections. This finding aligns with observations from other endemic
livestock diseases, where local spatial spread often sustains
transmission even when direct animal movements are restricted
\citep{Beninca2020, Boender2007}. Our result is also consistent with
regional evidence: \citet{Agren2016} identified Öland as a distinct
endemic hotspot and found that the presence of positive herds within a
5 km radius was strongly associated with seropositivity, whereas
animal purchases were not. Although their cross-sectional serological
approach differs from our longitudinal transmission model, both
studies converge on the importance of local transmission in sustaining
\textit{S.}~Dublin endemicity on Öland.

However, our findings contrast with \citet{Conrady2024}, who reported
that cattle movement activities were the strongest predictive factor
for \textit{S.}~Dublin infection, with twice the effect of local
transmission. Several methodological differences likely explain this
discrepancy. \citet{Conrady2024} employed social network analysis with
monthly aggregation of movements across 2010--2020, a method they
acknowledged might ``overrepresent connectivity'', potentially leading
to an overestimation of transmission effects. Furthermore, their
approach relied on logistic regression of observed infection statuses,
which does not explicitly account for observation error or the
temporal lag between infection and detection. In contrast, our study
utilises daily movement data and an SSM inferred via PMCMC. This
framework allows for the explicit estimation of latent infection
dynamics, disentangling the timing of movement introductions from
local transmission events while accounting for observation error. By
avoiding the temporal smoothing inherent in monthly aggregation and
directly inferring the underlying force of infection, our model
suggests that local spread plays a more substantial role than
previously estimated when relying on aggregated movement networks.

Building on these findings, our intervention simulations further
illuminate the relative importance of different transmission
pathways. Restricting movements of infected cattle alone was
insufficient to control transmission of \textit{S.}~Dublin, with the
effective reproduction number ($R_t^m$) remaining approximately at
unity and showing no evident downward trend. In contrast,
interventions targeting local spread resulted in a net reduction in
$R_t^d$, with the median eventually declining toward zero despite
initial fluctuations. This finding aligns with our observation that
local spread accounts for roughly 50\% of daily incident cases and
supports the metapopulation perspective that local transmission
between neighboring holdings is the primary driver of endemic
persistence in established clusters.  It is important to note that
these conclusions apply specifically to regions where the infection is
already endemic. In such settings, the ineffectiveness of movement
restrictions in isolation suggests that while animal movements are
critical for introducing infection to new holdings, they play a
secondary role in sustaining ongoing transmission within established
clusters.  Consequently, caution is warranted against increasing
movement intensity in these endemic areas, as this could increase the
relative contribution of movement-driven transmission and undermine
the gains from local control.  Therefore, control strategies that
prioritise reducing local transmission---through enhanced biosecurity,
spatial separation of holdings, or improved manure management---are
likely to be more effective than those focused solely on movement
restrictions, particularly in high-density areas like Öland where our
results suggest that local spread is the primary driver of
transmission dynamics.

The spatial configuration of farms plays a critical role in shaping
these dynamics, particularly through metapopulation mechanisms where
local clusters act as multipliers of infection.  \citet{Beninca2020}
demonstrated that in highly clustered populations, epidemic size is
maximised at intermediate transmission ranges that balance
within-cluster propagation and between-cluster dispersal.  Crucially,
once infection establishes in a high-density cluster, it can persist
even if transmission between clusters is limited, as local clusters
become self-sustaining reservoirs.  Similarly, \citet{Boender2007}
showed that the local reproduction number is directly proportional to
farm density, identifying critical density thresholds above which
epidemic spread becomes self-sustaining.  The density of farms on
Öland likely exceeds these critical thresholds, creating conditions
where local clusters function as persistent infection reservoirs,
consistent with our finding that local spread drives endemicity.  This
metapopulation perspective suggests that control strategies for
\textit{S.} Dublin should target both within-cluster biosecurity (to
reduce local reproduction numbers below unity) and between-cluster
transmission pathways (to prevent infection spread to new clusters),
as eliminating one route alone may be insufficient to disrupt the
overall transmission network.

Beyond transmission dynamics, we examined how diagnostic uncertainty
affects our estimates of \textit{S.}~Dublin prevalence. To assess the
robustness of our findings, we estimated model parameters under three
ELISA ODC\% cut-off scenarios and used these scenario-specific
estimates in posterior predictive simulations. This revealed that
diagnostic accuracy primarily influences observed holding-level
prevalence, which varied substantially across scenarios, while the
median within-holding prevalence of infected animals remained stable
at approximately 2--3\% and the overall prevalence across all holdings
at approximately 0.5\%. These stable within-herd estimates align
closely with empirical field data: the estimated 2--3\% prevalence of
actively infected animals matches the low faecal shedding rates
(typically $<$5\%) reported by \citet{Nielsen2013c} in endemically
infected Danish herds. Furthermore, the median proportion of
antibody-producing animals in our infected holdings ranged from 17\%
to 21\% across scenarios, a figure consistent with the seroprevalence
levels reported by \citet{Nielsen2013c} (varying from 19\% in calves
to 33\% in adult cows, with an overall average likely around
20--25\%).

Several limitations should be acknowledged when interpreting these
results. First, the model lacks explicit age stratification. While
\citet{Nielsen2013c} demonstrated that active shedding is concentrated
in calves, our model aggregates all animals into single
compartments. This simplification was necessary because the model was
parameterised using BTM data, which is collected only from lactating
cows. Consequently, we lacked the data required to independently
estimate transmission parameters for young stock. Although our
aggregated model successfully reproduced the net dynamics observed in
the field, it cannot capture age-specific transmission risks or the
specific role of calves in driving infection pressure. Second, while
herd size was included in the model, herd type (e.g., dairy versus
beef) was not fully distinguished, despite potential differences in
management practices that may influence \textit{S.}~Dublin risk.  Such
data were unavailable to stratify the population demographics and
animal movement networks by herd type.  Consequently, the model may
not fully capture the heterogeneity in transmission risks associated
with different management systems. Third, the model did not
distinguish between indoor housing versus pasture-based systems, which
could affect contact rates between animals and consequently
transmission dynamics. Animals on pasture may have different exposure
risks through environmental contamination or wildlife interactions
compared to housed animals. These limitations suggest that future
studies incorporating age-specific sampling---such as serological
surveys of young stock---would improve the precision of transmission
parameter estimates and enhance the targeting of control measures.

\section{Conclusion}
\label{conclusion}

While animal movements remain important for the introduction of
\textit{S.}~Dublin to new regions, the results suggest that local
spread is a dominant transmission route in maintaining endemicity,
particularly in geographically clustered regions like Öland. These
findings underscore the complexity of \textit{S.}~Dublin epidemiology
in spatially structured populations and demonstrate the value of using
a Bayesian simulation-based inference approach for explicitly
modelling latent infection dynamics and diagnostic
uncertainty. Effective \textit{S.}~Dublin control likely necessitates
a dual approach: enhancing surveillance to identify infected holdings,
while simultaneously implementing relevant biosecurity measures to
reduce local transmission and prevent the pathogen from persisting
within high-density clusters.

\section{CRediT authorship contribution statement}
\label{credit-authorship-contribution-statement}

\textbf{Stefan Widgren}: Conceptualization, Formal analysis, Funding
acquisition, Methodology, Software, Visualization, Writing - original
draft.  \textbf{Ivana R. Ewerlöf}: Data curation, Formal analysis,
Software, Visualization, Writing - review \& editing. \textbf{Arianna
  Comin}: Conceptualization, Formal analysis, Funding acquisition,
Writing - review \& editing.  \textbf{Robert Johansson}: Formal
analysis, Writing - review \& editing.  \textbf{Stefan Engblom}:
Conceptualization, Formal analysis, Funding acquisition, Methodology,
Software, Visualization, Writing - review \& editing.

\section{Declaration of Generative AI and AI-assisted technologies in
  the writing process}

During the preparation of this work the authors used Lumo in order to
improve the readability and language of the manuscript. After using
this tool/service, the authors reviewed and edited the content as
needed and take full responsibility for the content of the published
article.

\section{Declaration of competing interest}
\label{declaration-of-competing-interest}

The authors declare that they have no known competing financial
interests or personal relationships that could have appeared to
influence the work reported in this paper.

\section{Acknowledgments}
\label{acknowledgments}

This work was financially supported by Formas -- a Swedish Research
Council for Sustainable Development (grant number 2020-01887:
S. Widgren, A. Comin and S.  Engblom), (grant number 2021-02279:
S. Widgren and I. Ewerlöf) and by the Swedish Research Council (grant
number 2024-03949: S. Engblom).


\section{Data Availability}
\label{data-availability}

The data from the central database for bovine animals is confidential
and cannot be shared by the authors. The Swedish Board of Agriculture
owns the data, which has been shared with the Swedish Veterinary
Agency through an agreement.


\bibliographystyle{elsarticle-harv}
\bibliography{Salmonella-Dublin}

\end{document}